\documentclass{iau-JDSS}
\usepackage{graphicx}

\title[Modeling Dense Stellar Systems] 
{Modeling Dense Stellar Systems: Background}

\author[Piet Hut]   
{Piet Hut}

\affiliation{Institute for Advanced Study, Princeton, NJ 08540, USA
\break email: piet@ias.edu}

\pubyear{2006}
\volume{Volume 14}  
\pagerange{000--001}
\date{?? and in revised form ??}
\jname{Highlights of Astronomy, Volume 14}
\editors{K.A. van der Hucht, ed.}
\begin{document}

\maketitle

\begin{abstract}
I provide some background about recent efforts made in modeling dense
stellar systems, within the context of the MODEST initiative.  During
the last four years, we have seen more than fifteen MODEST workshops,
with an attendance between twenty and a hundred participants, and
topics ranging from very specialized discussions to rather general
overviews.
\end{abstract}

\firstsection 
\section{Dense Stellar Systems}

The study of star clusters, and of dense stellar systems in general,
has recently seen great progress, through observations as well as
simulations, as is evident from the papers in the proceedings of this
meeting, JD14.  The label `dense' is given when stars are close enough
that significant interactions between them occur on a time scale short
compared to the age of the stellar system.

A star forming region is
dense in this sense, because the contracting protostellar clouds have
a high probability to interact with each other during star formation.
Old open clusters are called dense when their age exceeds their
half-mass relaxation time.  Most globular clusters in our galaxy are
dense for that reason, and in addition, many globulars have a central
density high enough for physical collisions between stars to occur
frequently.

The most spectacular type of dense stellar system is that found in the
nucleus of most galaxies.  Our own Milky Way galaxy is no exception:
in the central parsec around the central supermassive black hole,
there are frequent collisions between the stars, which have a total
mass of a few million solar masses.

\section{Multi-Scale Simulations}

Twenty-five years ago, stellar dynamics was split up in a number of
different subfields that could be studied independently.  Planetary
dynamics, simulations of star forming regions, star cluster dynamics,
modeling of galactic nuclei, the study of interacting galaxies, and
cosmological simulations formed six different areas of research that
had rather little in common.

In contrast, all six areas are now firmly integrated.  In many cases,
it makes little sense to study only of these in isolation.  Starting
at the smallest scales, a detailed simulation of planet formation may
have to take into account the influence of neighboring stars within
the same star forming region.  Or looking from the largest scales,
that of cosmological simulations, the most detailed modeling efforts
resolve the encounters between individual galaxies; and in turn, a
detailed simulation of such an encounter shows how new dense star
clusters are formed in the process.

As a result, detailed simulations now routinely span multiple scales,
on which the same physical laws show rather different emerging
properties.  In stellar dynamics, relaxation effects
between stars can be ignored on galactic scales, yet are essential
in the more dense areas of star clusters and galactic nuclei.  And in
star forming regions, hydrodynamics shows quite different behaviors
on different scales.

\section{Multi-Physics Simulations}

The steady increase in computer power (\cite[Makino 2006]{m06}) has
made it possible to simulate multiple aspects of the physics of a
single system.  In a dense star cluster, we now can model the stellar
dynamical history of a modest cluster, as well as the stellar
evolution of each star.  In addition, we can also model the
hydrodynamical interactions that play a role when two or more stars
have a close encounter, possibly resulting in a collision.  Given that
computer speed has increased by a factor of a million in the last
thirty years, we can now follow the evolution of a million stars as
quickly as we could calculate the track of a single star in the mid
seventies.

As a consequence, the main bottlenecks in performing multi-scale,
multi-physics simulations are no longer related to hardware, but
rather to software limitations.  For example, none of the existing
legacy codes for stellar evolution can pass through all stages of
stellar evolution in a robust way, without human intervention.  One
priority is to develop simpler and more robust versions of all three
types of codes needed in the study of dense stellar systems, modeling
the stellar dynamics, evolution, and hydrodynamics.  Another priority
is to find ways to combine these codes in easy and realible ways.

\section{MODEST}

These two priorities have been the main aims of the MODEST initiative,
short for MOdeling DEnse STellar systems (see
http://www.manybody.org/modest.html).  Starting four years ago with
the first workshop in the American Museum for Natural History in New
York city (\cite[Hut et al. 2002]{h02}), we now hold a main workshop
each year, as well as a number of satellite meetings, typically once
every few months.  Most meetings have a few dozen participants,
though some of the yearly meetings have attracted a hundred
participants or more.  The topics of the workshops range from very
specialized discussions to rather general overviews.

\section{Frameworks}

Having robust versions of stellar dynamics, stellar evolution, and
hydrodynamics codes is not enough, if they cannot be coupled in
modular and flexible ways.  What is really needed is an umbrella
software environment, a {\it framework} that contains the `glue' to
connect different codes with each other.  In addition, such a
framework should contain user-friendly tools for visualization,
archiving, and comparisons with observations.

During the MODEST-6d workshop in Amsterdam in March 2006, we started a
first prototype framework for dense stellar systems, MUSE.  More
information can be found on the `projects' page of the MODEST home page,
mentioned above.  On the `workshops' page, you can find the schedule
of future frameworks-related workshops.  We plan to hold several such
workshops each year, to coordinate the ongoing development efforts.

\end{document}